\documentclass[epj,final]{svjour}
\usepackage{graphicx}           

\usepackage{axodraw}

\usepackage{cite}

\newcommand{\dsp}{\displaystyle}
\newcommand{\be}{\begin{equation}}
\newcommand{\ee}{\end{equation}}
\newcommand{\ba}{\begin{eqnarray}}
\newcommand{\ea}{\end{eqnarray}}

\newcommand{\ve}{\varepsilon}
\newcommand{\cO}{{\cal O}}

\newcommand{\Q}{{\cal Q}}

\begin{document}
\thispagestyle{empty}
\begin{onecolumn}
\begin{flushright}
\large
LU TP 04-40\\
hep-ph/0501163\\
December 2004
\end{flushright}
\vfill
\begin{center}
{\Large\bf Isospin Breaking in $K\to3\pi$ Decays III:\\[0.3cm] 
Bremsstrahlung and Fit to Experiment
}\\[1.5cm]

{\large \bf Johan Bijnens and Fredrik Borg}\\[1cm]
{\large Department of Theoretical Physics, Lund University\\[0.2cm]
S\"olvegatan 14A, S 22362 Lund, Sweden}

\vfill

{\bf Abstract}
\end{center}

We complete here our work on isospin violation in the $K\to3\pi$
system. We first calculate $K\to2\pi$ to the same order as we did
$K\to3\pi$ in papers I and II of this series. This adds effects of
order $G_{27}\,p^2 (m_u-m_d)$ and $G_{27}\,p^2 e^2$ to earlier work.
We calculate also the lowest order Bremsstrahlung contributions,
$K\to2\pi\gamma,3\pi\gamma$. With these and our earlier results
we perform a full fit to all available CP conserving data in the
$K\to2\pi,3\pi$ system including isospin violation effects.
We perform these fits under various input assumptions as well as test
the factorization and the vector dominance model for the weak NLO
low energy constants.

\vfill


\end{onecolumn}
\newpage
\setcounter{page}{0}

\title{Isospin Breaking in $K\to3\pi$ Decays III: 
Bremsstrahlung and Fit to Experiment\thanks{Supported in part by 
the European Union TMR
network, Contract No. HPRN-CT-2002-00311  (EURIDICE).}
}

\author{Johan Bijnens \and Fredrik Borg}
\institute{Department of Theoretical Physics, Lund University\\
S\"olvegatan 14A, S 22362 Lund, Sweden}

\abstract{
We complete here our work on isospin violation in the $K\to3\pi$
system. We first calculate $K\to2\pi$ to the same order as we did
$K\to3\pi$ in papers I and II of this series. This adds effects of
order $G_{27}\,p^2 (m_u-m_d)$ and $G_{27}\,p^2 e^2$ to earlier work.
We calculate also the lowest order Bremsstrahlung contributions,
$K\to2\pi\gamma,3\pi\gamma$. With these and our earlier results
we perform a full fit to all available CP conserving data in the
$K\to2\pi,3\pi$ system including isospin violation effects.
We perform these fits under various input assumptions as well as test
the factorization and the vector dominance model for the weak NLO
low energy constants.}


\maketitle

\section{Introduction}
\label{introduction}

Low-energy QCD is non-perturbative, which calls for alternative methods
of calculating processes including composite particles such as 
mesons and baryons. 
A method used to describe the interactions of the light pseudoscalar mesons 
($K,\pi,\eta$) is Chiral Perturbation Theory (ChPT). It was first 
presented by 
Weinberg, Gasser and Leutwyler \cite{Weinberg,GL1,GL2} and 
it has been very successful. Pedagogical introductions 
to ChPT can be found in \cite{chptlectures}. 
The theory can be
extended to also cover the weak interactions of the pseudoscalars, first
done in \cite{KMW1}. 

The first calculation of a kaon decaying into pions ($K\to2\pi,3\pi$) 
was presented in \cite{KMW2}, and reviews of other applications of ChPT to 
nonleptonic weak interactions can be found in \cite{chptweakreviews}. 
 
The details from \cite{KMW2} were lost, but a recalculation in the isospin 
limit of $K\to2\pi$ to next-to-leading order was made in \cite{BPP,BDP} and
of $K\to3\pi$ in \cite{BDP,GPS}.
In \cite{BDP} a full fit to all 
experimental data existing at the time was made, and it was found 
that the decay rates and 
linear slopes agreed well. However, a small discrepancy was
found in the quadratic slopes, and this can have several different origins.
It could be an 
experimental problem or it could have a theoretical origin. In the latter 
case the corrections to the amplitude calculated in \cite{BDP} are threefold:
strong isospin breaking, electromagnetic (EM) isospin breaking or higher order
corrections. 

In \cite{BB} the strong isospin and local electromagnetic corrections were
investigated and it was found that the inclusion of those led to changes of
a few percent in the amplitudes. The local electromagnetic part was
also calculated in \cite{GPS}, in full agreement with our result 
after sorting out some misprints in \cite{GPS}, corrected in
\cite{GPS2}. In \cite{BB2} the
radiative corrections were added, which means that the full effects of
isospin breaking were studied. This led to changes in the amplitudes
of order 5--10 percent. Note that the results in \cite{BB2} 
disagree numerically with the
results for $K^+\to\pi^0\pi^0\pi^+$ of \cite{nehme}.

To answer the question whether isospin 
breaking removes the problem of fitting the quadratic slopes, a new full
fit has to be done. That is the main result in this paper, and in this new
fit we also include new experimental data \cite{istra,kloe}. We also
present recalculations of the amplitudes $K\to2\pi$, $K\to2\pi\gamma$ and 
$K\to3\pi\gamma$, all calculated to next-to-leading order and including first 
order isospin breaking, i.e.\ we include contributions proportional to
$p^2$, $m^2$, $e^2$, $m_u-m_d$ (leading order), and 
$p^4,p^2\,m^2,m^4,p^2\,e^2$, $m^2\,e^2$, $p^2 (m_u-m_d)$ and $m^2 (m_u-m_d)$ 
(next-to-leading order). 
The corrections needed to be added to determine $\pi\pi$ scattering lengths
from $K\to3\pi$ \cite{cabibbo} are beyond the order calculated in this paper.

The outline of this paper is as follows. The next section describes isospin 
breaking in more detail. In section \ref{lagrangian} the basis of ChPT, 
the Chiral Lagrangians, are discussed. Section \ref{kinematics} 
specifies the decays and describes the relevant kinematics. The
divergences appearing when including photons are discussed in section 
\ref{isoIR}. In section \ref{isoanalyticres} the analytical results are 
discussed, section \ref{isonumres} contains the numerical results and the
last section contains the summary. 

\section{Isospin Breaking}
\label{striso}

Isospin symmetry is the SU(2) symmetry under the exchange of up-
and down-quarks. This symmetry is only exact in the approximation
that $m_u=m_d$ and electromagnetism is neglected, i.e.\ in the isospin limit.
Calculations are sometimes performed in the isospin limit since this is 
simpler and gives a good first estimate of the result.
However, to get a more accurate result one should include the effects 
from $m_u\neq m_d$ and electromagnetism, i.e.\ isospin breaking. 

The two different sources of isospin breaking give rise to different
effects. Strong isospin breaking, coming from $m_u\neq m_d$, 
include mixing between $\pi^0$ and
$\eta$. This mixing leads to changes in the formulas for both the 
physical masses of
$\pi^0$ and $\eta$ as well as the amplitude for any process involving either 
of the two. For a detailed discussion see \cite{strongiso}.

The other source, electromagnetic isospin breaking, coming from the
fact that the up- and the down-quarks are charged, implies 
interactions with photons. This means both the addition of new Lagrangians 
at each order, as well as the introduction of new diagrams including 
explicit photons.  

\section{The ChPT Lagrangians}
\label{lagrangian}

The basis of our ChPT calculation is the various Chiral Lagrangians of
relevant orders. We work to
leading order in $m_u-m_d$ and $e^2$ but next-to-leading order in
$p^2$ and $m^2$. For simplicity we call 
in the remainder terms of order 
$p^2$, $m^2$, $e^2$, $m_u-m_d$ leading order and terms 
of order $p^4$, $p^2\,m^2$, $m^4$,
$p^2\,e^2$, $m^2\,e^2$, $p^2 (m_u-m_d)$ and $m^2 (m_u-m_d)$
next-to-leading order.
 
\subsection{Leading Order}
\label{leading}

The lowest order Chiral Lagrangian is divided in three parts
\be
{\cal L}_{2}={\cal L}_{S2}+{\cal L}_{W2}+{\cal L}_{E2},
\ee
where ${\cal L}_{S2}$ refers to the strong $\Delta S = 0$ part, 
${\cal L}_{W2}$ the weak 
$\Delta S = \pm 1$ part, and ${\cal L}_{E2}$ the strong-electromagnetic and 
weak-electromagnetic parts combined. For the strong part we have \cite{GL1}
\be
\label{L2S}
{\cal L}_{S2} = \frac{F_0^2}{4} \; \langle u_\mu u^\mu + \chi_+\rangle\,.
\ee
Here $\langle A\rangle$ stands for the flavour trace of the matrix $A$,
and $F_0$ is the pion decay constant in the chiral limit.
We define the matrices $u_\mu$, $u$ and $\chi_\pm$ as
\be
u_\mu = i u^\dagger\, D_\mu U\, u^\dagger = u_\mu^\dagger\,,\quad u^2 = U\,,
\quad \chi_\pm = u^\dagger \chi u^\dagger \pm u \chi^\dagger u\,,
\ee
where the special unitary matrix $U$ contains the Goldstone boson fields
\ba
U &=& \exp\left(\frac{i\sqrt{2}}{F_0}M\right)\,,\,\,
\nonumber\\
M &=&\left(\begin{array}{ccc}
\frac{1}{\sqrt{2}}\pi_3+\frac{1}{\sqrt{6}}\eta_8 & \pi^+ & K^+\\
\pi^- & \frac{-1}{\sqrt{2}}\pi_3+\frac{1}{\sqrt{6}}\eta_8 & K^0\\
K^- & \overline{K^0} & \frac{-2}{\sqrt{6}}\eta_8
         \end{array}\right)\,.
\ea
We use the formalism of the external field method \cite{GL1}, 
and to include photons we set
\be
\chi = 2 B_0
\left(\begin{array}{ccc}m_u &  & \\ & m_d & \\ & & m_s\end{array}\right)\, 
\ee
and
\be
D_\mu U = \partial_\mu U - i e\, Q A_\mu U
 - i e\, U Q A_\mu,
\ee
where $A_\mu$ is the photon field and 
\begin{equation} \label{defZ}
Q = \left( \begin{array}{ccc} 2/3 & &\\ & -1/3 & \\ & & -1/3
\end{array}
\right).
\ee
The quadratic terms in (\ref{L2S}) are diagonalized by a rotation
\ba
\pi^0 &=& \pi_3\cos\epsilon + \eta_8\sin\epsilon\,
\nonumber\\
\eta &=& -\pi_3\sin\epsilon + \eta_8\cos\epsilon\,,
\ea
where the lowest order mixing angle $\epsilon$ satisfies
\be
\tan(2\epsilon) = \sqrt{3}\frac{m_d-m_u}{2\,m_s-m_u-m_d}\,.
\ee

The weak, $\Delta S=1$, part of the Lagrangian has the form \cite{Cronin}
\ba
\label{LW2}
{\cal L}_{W2} &=& C \, F_0^4 \, 
\Bigg[ G_8 \langle \Delta_{32} u_\mu u^\mu \rangle +
G_8' \langle\Delta_{32} \chi_+ \rangle 
\nonumber\\&&
+G_{27} t^{ij,kl} \, \langle \Delta_{ij } u_\mu \rangle
\langle\Delta_{kl} u^\mu \rangle \Bigg]
+ \mbox{ h.c.}\,.\nonumber
\ea
The tensor $t^{ij,kl}$ has as nonzero components
\ba
\label{deft}
t^{21,13} =
t^{13,21} = \frac{1}{3} \,, && \, 
t^{22,23}=t^{23,22}=-\frac{1}{6} \, , \nonumber \\
t^{23,33}=t^{33,23}=-\frac{1}{6} \,, && \, 
t^{23,11} =t^{11,23}=\frac{1}{3}\,,
\ea
and the matrix $\Delta_{ij}$ is defined as
\be
\Delta_{ij} \equiv u \lambda_{ij} u^\dagger\,,\quad
\left(\lambda_{ij}\right)_{ab} \equiv \delta_{ia} \, \delta_{jb}\,.
\ee
The coefficient $C$ is defined such that in the chiral
and large $N_c$ limits $G_8 = G_{27} =1$,
\be
C= -\frac{3}{5} \, \frac{G_F}{\sqrt 2} V_{ud} \, V_{us}^* = 
-1.06\cdot 10^{-6}\,\, \mathrm{GeV}^{-2}\, .
\ee
Finally, the remaining electromagnetic part, relevant for this calculation,
 looks like (see e.g. \cite{EMEcker})
\be
\label{LE2}
{\cal L}_{E2} = 
e^2 F_0^4 Z \langle {\cal Q}_L {\cal Q}_R\rangle +
e^2 F_0^4 \langle \Upsilon {\cal Q}_R\rangle\,,
\ee
where the weak-electromagnetic term is multiplied by a 
constant $G_E$ \mbox{($g_{\rm ewk}G_8$ in \cite{EMEcker})},
\begin{equation} \label{gewk}
\Upsilon = G_E\, F_0^2 \Delta_{32} + {\rm h.c.}  \, 
\ee
and 
\begin{equation}
\Q_L = uQu^\dagger\,,\,\,\, \Q_R = u^\dagger Qu \, .
\ee

\subsection{Next-to-leading Order}
\label{ntleading}

Chiral Perturbation Theory is a non-renormalizable theory. This means 
that new terms have to be added at 
each order to compensate for the divergences coming from loop-diagrams. 
Thus
the Lagrangians increase in size for every new order and the number of 
free parameters rises as well.
At next-to-leading order the Lagrangian is split in four parts which, 
in obvious notation, are 
\be
{\cal L}_{4}={\cal L}_{S4}+{\cal L}_{W4}+{\cal L}_{S2E2}+{\cal L}_{W2E2}(G_8)
\,.
\ee
The notation $(G_8)$ indicates that here only the dominant
$G_8$-part is included in the Lagrangian and therefore in the calculation.

The Lagrangians of next-to-leading order are quite large and we will not 
write them explicitly 
here since they can be found in many places 
\cite{GL1,radiative,KMW1,Esposito,EKW,EMEcker,urech}. 
For a list of all the pieces relevant for this specific calculation 
see \cite{BB,BB2}.
Note however that one contributing term was forgotten 
when writing ${\cal L}_{S2E2}$ in \cite{BB}, namely
\ba
&\dsp
-i\,e^2\,F_0^2\,K_{12}\,\langle(\widehat \nabla_\mu \Q_L \Q_L -
\Q_L \widehat \nabla_\mu \Q_L &
\nonumber\\
&\dsp  - \widehat \nabla_\mu \Q_R \Q_R +
\Q_R \widehat \nabla_\mu \Q_R)\, u^\mu\rangle\,,&
\ea
where
\ba
\widehat \nabla_\mu \Q_L &=& \nabla_\mu \Q_L + \frac{i}{2} [u_\mu,\Q_L] =
u D_\mu Q_L u^\dagger , \nonumber \\
\widehat \nabla_\mu \Q_R &=& \nabla_\mu \Q_R - \frac{i}{2} [u_\mu,\Q_R] =
u^\dagger D_\mu Q_R u\,.
\ea
It contributes to the calculation of the decay constants, $F_{\pi^+}$ and 
$F_{K^+}$. It only contributes to the amplitudes of $K\to2\pi$ and $
K\to3\pi$ via the rewriting of the lowest order in terms of $F_{\pi^+}$
and $F_{K^+}$ rather than $F_0$.

\subsubsection{Ultraviolet Divergences}

The processes $K\to 2\pi$ and $K\to 3\pi$ receives 
higher-order contributions from diagrams that contain loops. 
The study of these diagrams is complicated
by the fact that they need to be precisely defined. The loop-diagrams involve
an integration over the loop-momentum $Q$, and the 
integrals are divergent in the ultraviolet region, i.e.\ when  
$Q\rightarrow\infty$.
These ultraviolet divergences are canceled by
replacing the coefficients, $X_i$, in the next-to-leading order Lagrangians by
the renormalized coefficients, $X_i^r$, and a subtraction part. 
See \cite{BDP,BB} and references therein.

\section{Kinematics}
\label{kinematics}

\subsection{$K\to2\pi$ and $K\to2\pi\gamma$}

In the limit of CP-conservation, there are three different decays of the type
$K\to2\pi$ ($K^-$ decays are not treated separately since they are counterparts
to the $K^+$ decays):
\ba
\label{defdecaysK2P}
K_S(k)&\to&\pi^0(p_1)\,\pi^0(p_2)\,,\quad [A^S_{00}]\,,\nonumber\\
K_S(k)&\to&\pi^+(p_1)\,\pi^-(p_2)\,,\quad [A^S_{+-}]\,,\nonumber\\
K^+(k)&\to&\pi^+(p_1)\,\pi^0(p_2)\,,\quad [A_{+0}],
\ea
where we have indicated the four-momentum defined for each particle
and the symbol used for the amplitude.
With an external photon it changes to:
\ba
\label{defdecaysK2PG}
K_S(k)&\to&\pi^0(p_1)\,\pi^0(p_2)\,\gamma\,(q)\,,\quad [A^S_{00\gamma}]\,,\nonumber\\
K_S(k)&\to&\pi^+(p_1)\,\pi^-(p_2)\,\gamma\,(q)\,,\quad [A^S_{+-\gamma}]\,,\nonumber\\
K^+(k)&\to&\pi^+(p_1)\,\pi^0(p_2)\,\gamma\,(q)\,,\quad [A_{+0\gamma}]\,.
\ea
The kinematics for $K\to2\pi\gamma$ is treated using
\be
r_0 \equiv -k\cdot q\,,\quad
r_1 \equiv p_1\cdot q\,,\quad
r_2 \equiv p_2\cdot q\,,\quad
\ee
where
\be
r_0+r_1+r_2 = 0\,.
\ee

\subsection{$K\to3\pi$ and $K\to3\pi\gamma$}

For the corresponding process $K\to3\pi$, there are five different decays:
\ba
\label{defdecaysK3P}
K_L(k)&\to&\pi^0(p_1)\,\pi^0(p_2)\,\pi^0(p_3)\,,\quad [A^L_{000}]\,,\nonumber\\
K_L(k)&\to&\pi^+(p_1)\,\pi^-(p_2)\,\pi^0(p_3)\,,\quad [A^L_{+-0}]\,,\nonumber\\
K_S(k)&\to&\pi^+(p_1)\,\pi^-(p_2)\,\pi^0(p_3)\,,\quad [A^S_{+-0}]\,,\nonumber\\
K^+(k)&\to&\pi^0(p_1)\,\pi^0(p_2)\,\pi^+(p_3)\,,\quad [A_{00+}]\,,\nonumber\\
K^+(k)&\to&\pi^+(p_1)\,\pi^+(p_2)\,\pi^-(p_3)\,,\quad [A_{++-}]\,,
\ea
and here the variables are
\be
s_1 \equiv \left(k-p_1\right)^2\,,\quad
s_2 \equiv \left(k-p_2\right)^2\,,\quad
s_3 \equiv \left(k-p_3\right)^2\,,
\ee 
where
\be
s_1+s_2+s_3 = k^2+p_1^2+p_2^2+p_3^2\,.
\ee
The amplitudes are expanded in terms of the Dalitz plot variables
$x$ and $y$ defined as
\be
\label{defsi}
y =\frac{ s_3-s_0}{m_{\pi^+}^2}\,,\quad
x =\frac{ s_2-s_1}{m_{\pi^+}^2}\,,\quad
s_0 = \frac{1}{3}\left(s_1+s_2+s_3\right)\,.
\ee
With an external photon the decays are:
\ba
\label{defdecaysK3PG}
K_L(k)&\to&\pi^0(p_1)\,\pi^0(p_2)\,\pi^0(p_3)\,\gamma\,(q)\,,\quad [A^L_{000\gamma}]\,,\nonumber\\
K_L(k)&\to&\pi^+(p_1)\,\pi^-(p_2)\,\pi^0(p_3)\,\gamma\,(q)\,,\quad [A^L_{+-0\gamma}]\,,\nonumber\\
K_S(k)&\to&\pi^+(p_1)\,\pi^-(p_2)\,\pi^0(p_3)\,\gamma\,(q)\,,\quad [A^S_{+-0\gamma}]\,,\nonumber\\
K^+(k)&\to&\pi^0(p_1)\,\pi^0(p_2)\,\pi^+(p_3)\,\gamma\,(q)\,,\quad [A_{00+\gamma}]\,,\nonumber\\
K^+(k)&\to&\pi^+(p_1)\,\pi^+(p_2)\,\pi^-(p_3)\,\gamma\,(q)\,,\quad [A_{++-\gamma}]\,,
\ea
where the kinematics is treated using
\be
s_{1\gamma} \equiv \left(k-p_1\right)^2,\,
s_{2\gamma} \equiv \left(k-p_2\right)^2,\,
s_{3\gamma} \equiv \left(k-p_3\right)^2,
\ee
\be
t_0 \equiv -k\cdot q\,,\quad 
t_1 \equiv p_1\cdot q\,,\quad 
t_2 \equiv p_2\cdot q\,,\quad 
t_3 \equiv p_3\cdot q\,,
\ee
where 
\be
t_0 + t_1 + t_2 + t_3 = 0\,
\ee
and
\be
s_{1\gamma}+s_{2\gamma}+s_{3\gamma} = k^2+p_1^2+p_2^2+p_3^2-2\,t_0\,.
\ee
\section{Infrared Divergences}
\label{isoIR}

Beside the ultraviolet divergences which are removed by
renormalization of the higher order coefficients, diagrams including 
photons in the loops contain infrared (IR)
divergences. These
infinities come from the $Q\rightarrow 0$ end of the loop-momentum 
integrals. They are handled by including also the 
Brems\-strahlung process, where a real photon
is radiated off one of the charged mesons.
It is only the sum of the virtual loop corrections and the real Bremsstrahlung
which is physically significant and thus needs to be well defined.

We regulate the IR divergences in both the virtual photon loops and the real
emission with a photon mass $m_\gamma$ and keep only the singular terms
plus those that do not vanish in the limit $m_\gamma\to 0$. We include the 
real Bremsstrahlung for photon energies up to a cut-off $\omega$ and
treat it in the soft photon approximation.

The exact
form of the amplitude squared for the Brems\-strahlung
depends on which specific amplitude that is 
being calculated. For a detailed presentation of the calculation and 
resulting expressions for $K\to3\pi$ see \cite{BB2}. The corresponding
amplitudes for $K\to2\pi$ are
\be
|A^S_{00}|^2_{BS} = 0 \,,
\ee
\be
|A^S_{+-}|^2_{BS} = -|A^S_{+-}|^2_{LO} \,\frac{e^2}{4\pi^2}\,\left[\log 
  \frac{\omega^2}{m_\gamma^2} - I_{IR}(m_\pi^2,m_\pi^2,m_K^2)\right]
\ee
\be
|A_{+0}|^2_{BS} = -|A_{+0}|^2_{LO} \,\frac{e^2}{4\pi^2}\,\left[\log 
  \frac{\omega^2}{m_\gamma^2} - I_{IR}(m_\pi^2,m_K^2,m_\pi^2)\right]
\ee
where
\be
I_{IR}(m_1^2,m_2^2,m_3^2) \equiv -\frac{x_s}{4\pi^2} \frac{m_3^2-m_1^2-m_2^2}
{m_1 m_2 (1-x_s)}
\log x_s \log \frac{\omega^2}
{m_\gamma^2}. 
\ee
When using these expressions, the divergences from the photon 
loops cancel exactly. 

A similar problem shows up in the definition of the decay constants, since
we normalize the lowest order with $F_{\pi^+}$ and $F_{K^+}$. See \cite{BB2}
for details.

\section{Analytical Results}
\label{isoanalyticres}

\subsection{$K\to 2\pi$}

The most complete work on isospin violation in $K\to 2\pi$ is 
in \cite{k2piisofull},
earlier work can be found in \cite{K2PIiso}.

\subsubsection{Lowest Order}

There is only one diagram contributing to the decay $K\to 2\pi$ at lowest 
order, see top left in Fig.~\ref{fig:K2P}, and the resulting 
amplitudes are also
quite simple. To first order in isospin they can be written
\ba
A^S_{00} &=& i\,F\,(G_8-G_{27}) \\ && \Bigg(
            4\,\frac{\sin\epsilon}{\sqrt{3}}\,(m_\pi^2-m_K^2)
          - 2\,m_{\pi^0}^2
          + 2\,m_{K^0}^2
          \Bigg)\,,
\ea
\ba
A^S_{+-} &=& i\,F\,G_8 \, (
          - 2\,m_{\pi^+}^2
          + 2\,m_{K^0}^2
          )
\nonumber\\&&
       + i\,F\,G_{27} \, \left(
          - \frac{4}{3}\,m_{\pi^+}^2
          + \frac{4}{3}\,m_{K^+}^2
          \right)
   - 2\,i\,F^3\,e^2\,G_E \,,
\nonumber \\    &&
\ea
\ba
A_{+0}&=&i\,F\,G_8 \, \left(
          - 2\,\frac{\sin\epsilon}{\sqrt{3}}\,m_\pi^2
          + 2\,\frac{\sin\epsilon}{\sqrt{3}}\,m_K^2
          - m_{\pi^+}^2
          + m_{\pi^0}^2
          \right)
 \nonumber \\&&
       + i\,F\,G_{27} \, \Bigg(
          - 3\,\frac{\sin\epsilon}{\sqrt{3}}\,m_\pi^2
          + 3\,\frac{\sin\epsilon}{\sqrt{3}}\,m_K^2
          - 7/3\,m_{\pi^+}^2
\nonumber \\&&
          + 2/3\,m_{\pi^0}^2
          + 5/3\,m_{K^+}^2
          \Bigg)
- i\,F^3\,e^2\,G_E\,.
\ea
See section~\ref{sec:SEInput} for a discussion of the masses used.
\begin{figure}
\begin{center}
\includegraphics[width=0.45\textwidth]{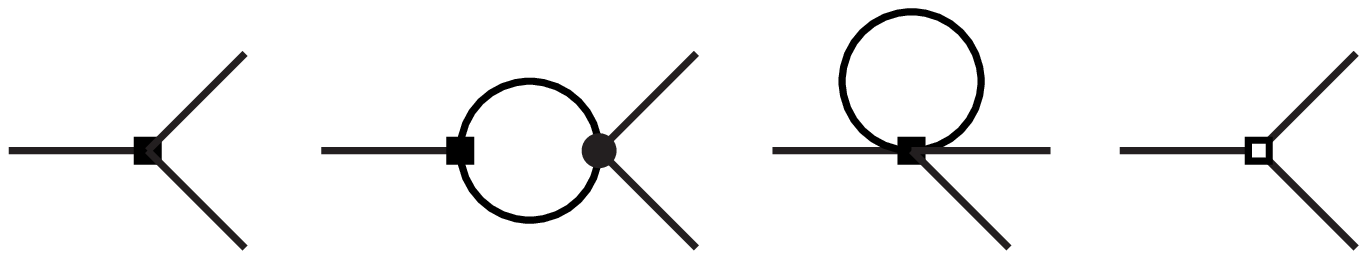}
\\
\includegraphics[width=0.45\textwidth]{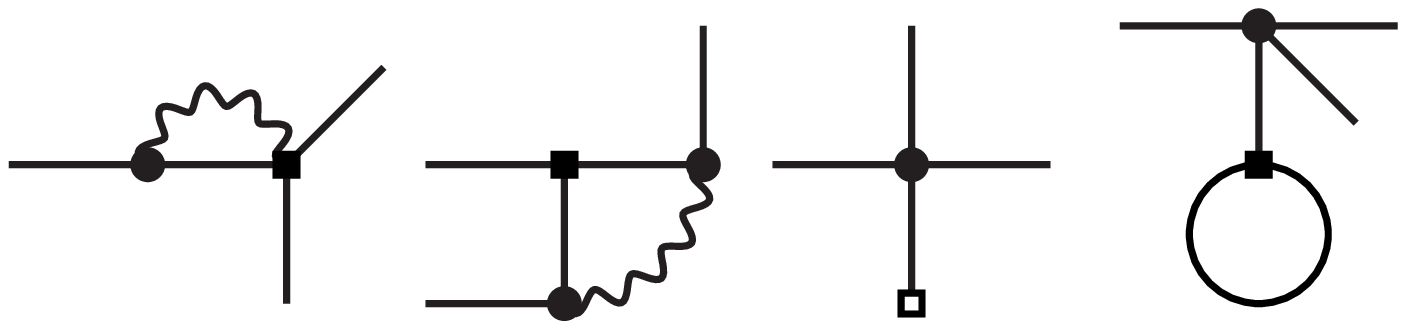}
\end{center}
\caption{The diagrams for $K\to2\pi$. An open square is a vertex
from ${\cal L}_{W4}$ or ${\cal L}_{W2E2}$,
a filled square a vertex
from ${\cal L}_{W2}$ or ${\cal L}_{E2}\, (\Delta S = 1)$ and a filled 
circle a vertex 
from ${\cal L}_{S2}$ or ${\cal L}_{E2}\, (\Delta S = 0)$. A straight 
line is a pseudoscalar meson and a wiggly line a photon.
\label{fig:K2P}}
\end{figure}

\subsubsection{Next-to-leading Order}

There are seven more diagrams contributing to next-to-leading order, 
see Fig.~\ref{fig:K2P}. The resulting amplitudes are long, and we decided
to not include them here but instead make them available for download
\cite{formulas}. 
Note that we have also included contributions proportional to $G_{27}$, 
not included in \cite{k2piisofull}. These are included for consistency between
the $K\to 2\pi (\gamma)$ and $K\to 3\pi (\gamma)$ calculations.

\subsection{$K\to 2\pi \gamma$}

The amplitudes for the processes $K\to 2\pi \gamma$ have been calculated
before. Here we only need the lowest order contribution to be consistent
with the $K\to2\pi$ calculation.
This we  recalculated and the starting point
is the two diagrams that contribute to the process, shown in
Fig.~\ref{fig:K2PG}.

\begin{figure}
\begin{center}
\setlength{\unitlength}{1pt}
\includegraphics[width=0.30\textwidth]{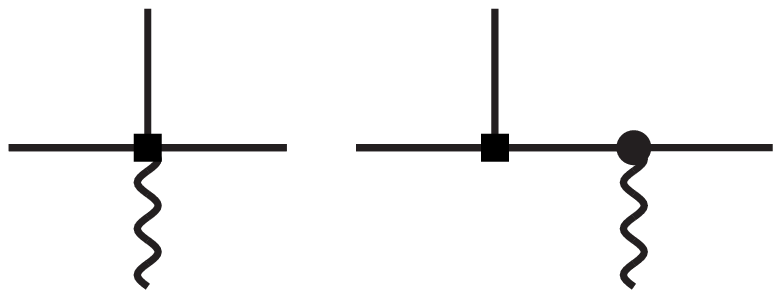}
\end{center}
\caption{The diagrams for $K\to2\pi\gamma$. A square is a 
weak vertex, a circle a strong vertex, a straight line a pseudoscalar
meson and a wiggly line a photon.
\label{fig:K2PG}}
\end{figure}
Since photons at this order
only couple to charged particles, there are only two 
different amplitudes and the results are
\ba
A^S_{+-\gamma}& =& e\,F\,\left(G_8 + \frac{2}{3}\,G_{27}\right)\,
(m_K^2-m_\pi^2)
\nonumber \\ && 
\left[-k.\varepsilon\,\left(\frac{1}{r_1}
        -\frac{1}{r_2}\right)
  +(p_2.\varepsilon\,-p_1.\varepsilon)\left(\frac{1}{r_1}+\frac{1}{r_2}\right)
   \right]\,,
\nonumber\\&&
\ea 
and
\ba
A_{+0\gamma}& =& e\,F\,\frac{5}{6}\,G_{27}\,(m_K^2-m_\pi^2)\,
\left[-k.\varepsilon\,\left(\frac{1}{r_0}
         +\frac{1}{r_1}\right)\right.
\nonumber \\  &&
   \left.+p_1.\varepsilon\,\left(-\frac{1}{r_0}-\frac{1}{r_1}\right)
   +p_2.\varepsilon\,\left(-\frac{1}{r_0}+\frac{1}{r_1}\right)\right]\,.
\ea
These amplitudes  
can be decomposed into an electric and a magnetic part:
\be
A(K \to 2 \pi \gamma) = e \,\ve^\mu(k)\,(E_\mu + 
\ve_{\mu\nu\rho\sigma} M^{\nu\rho\sigma})\,,
\ee
but at lowest order the magnetic amplitude $M^{\nu \rho\sigma}$ 
vanishes since
there is no $\ve_{\mu\nu\rho\sigma}$ tensor in the corresponding lowest 
order Lagrangian. 

The electric
amplitude, on the other hand, is completely determined by the corresponding
non-radiative amplitude via Low's theorem \cite{Low58}.

\subsection{$K\to 3\pi \gamma$}

The decay $K\to 3\pi \gamma$ is discussed in detail in \cite{K3PG}. 
We only need the lowest order amplitude for consistency with the calculation
of $K\to3\pi$.
We have 
calculated the four different amplitudes using Chiral Perturbation Theory, and 
checked that they agree with Low's theorem. The calculation is
based on seven diagrams, see Fig.~\ref{fig:K3PG}.
\begin{figure}
\begin{center}
\includegraphics[width=0.45\textwidth]{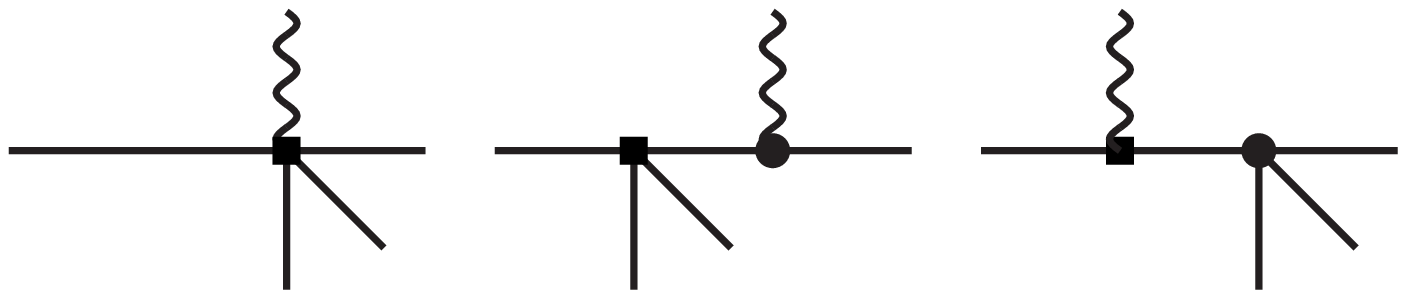}
\\[0.5cm]
\includegraphics[width=0.45\textwidth]{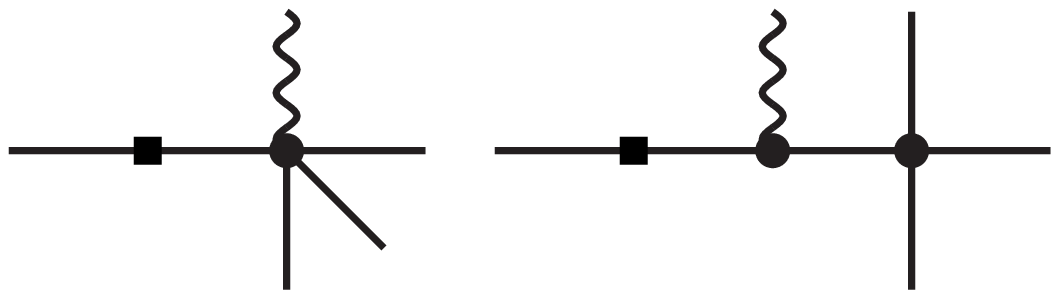}
\\[0.5cm]
\includegraphics[width=0.45\textwidth]{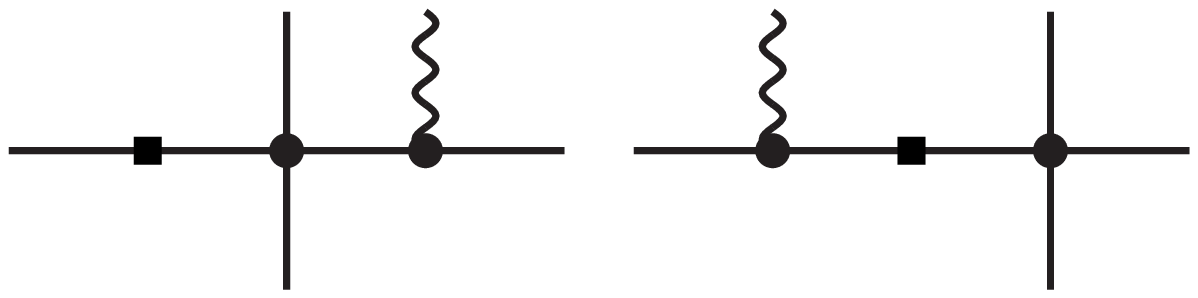}
\end{center}
\caption{The diagrams for $K\to3\pi\gamma$. A square is a 
weak vertex, a circle a strong vertex, a straight line a pseudoscalar
meson and a wiggly line a photon.
\label{fig:K3PG}}
\end{figure}
The four amplitudes are
\ba
A^L_{+-0\gamma}& = & i\,e\,
   \left[\frac{G_8-G_{27}}{3}\,m_K^2+\right.
\nonumber\\&&
\left(-\frac{G_8}{3}-\frac{G_{27}}{6}\,
          \frac{-3\,m_K^2+8\,m_\pi^2}{m_\pi^2-m_K^2}\right) \times
\nonumber\\
    &&        (-3\,(s_{3\gamma}-2\,t_1-2\,t_2)
 +m_K^2+3\,m_\pi^2) \times
\nonumber\\&&
   \,\left(\frac{p_1\cdot\ve}{t_1}-\frac{p_2\cdot\ve}{t_2}\right)\nonumber\\
  &&    -6\,\left(-\frac{G_8}{3}-\frac{G_{27}}{6}\,
          \frac{-3\,m_K^2+8\,m_\pi^2}{m_\pi^2-m_K^2}\right)\times \nonumber\\
&&   (t_2+t_1)\,\left(\frac{p_1\cdot\ve}{t_1}-\frac{p_2\cdot\ve}{t_2}\right)
          \Bigg]\,,
\ea
\ba
A^S_{+-0\gamma}& =& i\,e\,\frac{5}{6}\,
       \left(-\frac{G_{27}}{m_\pi^2-m_K^2}\right)\,
        (2\,m_\pi^2-3\,m_K^2)\times
\nonumber\\ &&
   \left[  (s_{1\gamma}-s_{2\gamma})\,
     \left(\frac{p_1\cdot\ve}{t_1}-\frac{p_2\cdot\ve}{t_2}\right)\right.  
\nonumber\\&&
+2\,t_0\,\left(\frac{p_1\cdot\ve}{t_1}+\frac{k\cdot\ve}{t_0}\right)
\nonumber\\
&&      
      +2\,t_0\,\left(\frac{p_2\cdot\ve}{t_2}+\frac{k\cdot\ve}{t_0}\right) 
  \Bigg]\,,
\ea
\ba
A_{00+\gamma}& =& i\,e\,
   \left[ -\frac{G_8}{2}\,(m_K^2+m_\pi^2)-\frac{G_{27}}{m_\pi^2-m_K^2}\,
          (5/3\,m_\pi^2\,m_K^2 \right.
\nonumber\\ &&
-13/6\,m_\pi^4          +1/2\,m_K^4)
  -\left(\frac{G_8}{2}+\frac{G_{27}}{m_\pi^2-m_K^2}\times \right.
\nonumber\\ &&
(7/6\,m_\pi^2-17/6\,m_K^2)
            \Big)\,\times \nonumber\\
&&
   (2\,(s_{3\gamma}-2\,t_1-2\,t_2)-m_K^2-3\,m_\pi^2)
\nonumber\\ &&
      -\frac{G_{27}}{m_\pi^2-m_K^2}\,5/6\,(2\,m_\pi^2-3\,m_K^2)\times 
\nonumber\\
&&
 (-(s_{3\gamma}-2\,t_1-2\,t_2)+m_K^2+3\,m_\pi^2)\Bigg]\times
\nonumber\\ &&
           \,\left(\frac{k\cdot\ve}{t_0}+\frac{p_3\cdot\ve}{t_3}\right)
\ea
\ba
A_{++-\gamma}& =& i\,e\,
   \Bigg[\Bigg(-G_8\,(m_\pi^2+m_K^2)+G_{27}\,(m_K^2+13/3\,m_\pi^2)
\nonumber\\ && 
+     (G_8-13/3\,G_{27}) (s_{3\gamma}-2\,t_1-2\,t_2) \Bigg) \times
\nonumber\\
&&
        \left(\frac{k\cdot\ve}{t_0}+\frac{p_1\cdot\ve}{t_1}
          +\frac{p_2\cdot\ve}{t_2}-\frac{p_3\cdot\ve}{t_3}\right)
\nonumber\\ &&
    +(2 G_8-26/3\,G_{27})\,
         (t_2-t_1)\left(\frac{p_1\cdot\ve}{t_1}-
             \frac{p_2\cdot\ve}{t_2}\!\right)\!\Bigg].
\nonumber\\ &&
\ea
Once again, these amplitudes  
can be decomposed into an electric and a magnetic part:
\be
A(K \to 3 \pi \gamma) = e \,\ve^\mu(k)\,(E_\mu + 
\ve_{\mu\nu\rho\sigma} M^{\nu\rho\sigma})\,,
\ee
and the magnetic amplitude $M^{\nu \rho\sigma}$ vanishes for the reasons
given above.  

The electric
amplitude at this order is again completely determined by the corresponding
non--radiative amplitude $A(s,\nu)$ via Low's theorem \cite{Low58,K3PG}:
\ba
\label{Low}
E^\mu &=& A(s,\nu)\, \Sigma^\mu \nonumber \\
&& \mbox{} + 2 \frac{\partial A(s,\nu)}{\partial s} \Lambda^\mu_{12} +
\frac{\partial A(s,\nu)}{\partial \nu} (\Lambda^\mu_{14} -
\Lambda^\mu_{24}) \nonumber \\
&& \mbox{} + \cO(k)
\ea
with (the meson charges in units of $e$ are denoted $q_i$, with
$\sum_{i=1}^4 q_i = 0$)
\ba
\label{defs}
s &=& (p_1+p_2)^2 \nonumber \\
\nu &=& k\cdot(p_1-p_2) \nonumber \\
\Sigma^\mu &=& \sum_{i=1}^4 \frac{q_i p_i^\mu}{t_i} \nonumber \\
\Lambda^\mu_{ij} &=& \Lambda^\mu_{ji} = (q_i t_j - q_j t_i) D^\mu_{ij}
\nonumber \\
D^\mu_{ij} &=& - D^\mu_{ji} = \frac{p_i^\mu}{t_i} - \frac{p_j^\mu}{t_j}~.
\ea
Since there are no terms of $\cO(k)$ at lowest order in the chiral expansion, 
the leading--order electric amplitude is completely determined by the 
explicit terms in (\ref{Low}). 

\section{Numerical Results}
\label{isonumres}

\subsection{Experimental Data and Input}

For the numerical studies we use the input given in 
Table~\ref{tab:isoInput}.
\begin{table}
\begin{center}
\begin{tabular}{|c|r||c|r|}
\hline
$G_E$   & $-0.4$ &$L_1^r$ & $0.38\cdot10^{-3}$  \\
$\sin\epsilon$ & $1.19\cdot 10^{-2}$& $L_2^r$ & $1.59\cdot10^{-3}$  \\
$Z$ & $0.805$ & $L_3^r$ & $-2.91\cdot10^{-3}$  \\
$F_\pi$&$0.0924$ GeV& $L_4^r$ & $0$  \\
$F_K$&$0.113$ GeV&$L_5^r$ & $1.46\cdot10^{-3}$ \\
$N_{14}$&$-10.4\cdot 10^{-3}$& $L_6^r$ & $0$  \\
$N_{15}$&$5.95\cdot 10^{-3}$&$L_7^r$ & $-0.49\cdot10^{-3}$  \\
$D_{13}$&$0$& $L_8^r$ & $1.0\cdot10^{-3}$  \\ 
$D_{15}$&$0$&$L_9^r$& $7.0\cdot10^{-3}$  \\
\hline
\end{tabular}
\end{center}
\caption{The various input values used, LECs given at $\mu=0.77$ GeV.
\label{tab:isoInput}}
\end{table}
\subsubsection{Strong and Electromagnetic Input}
\label{sec:SEInput}

There are different ways to treat the masses, especially in the isospin limit 
case. In \cite{BDP} the masses used in the phase space were obtained
from the physical masses
occurring in the decays. However, in the amplitudes the physical
mass of the kaon involved in the process was used and the pion 
mass was given by $m_{\pi}^2=\frac{1}{3}\sum_{i=1,3} m_{\pi^i}^2$
with $i=1,2,3$ being the three pions participating in the reaction. 
This allowed for the correct kinematical relation 
$s_1+s_2+s_3= m_K^2+3m_\pi^2$ to be satisfied while having the isospin
limit in the amplitude but the physical masses in the phase space.
The results in \cite{BDP} were obtained with the physical mass for the
eta.  Results with the Gell-Mann-Okubo (GMO) relation for the
eta mass in the loops gave small
changes within the general errors given in \cite{BDP}. 

In the decays here, we work to first order in isospin breaking. We have
rewritten explicit factors of $m_u-m_d$ in terms of $\sin\epsilon$ according
to
\be
m_u-m_d = -\frac{1}{\sqrt{3}}\,\, (2\,m_s-m_u-m_d)\,
\sin\epsilon \,.
\ee
In  general we use the physical masses of pions and kaons in the loops but
as soon as
a factor of $\sin\epsilon$ or $e^2$ is present we use a common kaon and
a common pion mass. This simplifies the analytical formulae enormously.
The kaon mass chosen is the mass from the kaon
in the decay and the pion mass used is $3 m_\pi^2 = \sum_i m_{\pi^i}^2$
with $i=1,2,3$ the three pions in the final state, i.e.\ the mass we used
in the isospin limit case. For the eta mass we use in general the
physical mass in the loop integrals.
We have used the GMO mass relation
with isospin violation included,
\be
m_\eta^2 = \frac{2}{3}\, (m_{K^+}^2+m_{K^0}^2-m_{\pi^+}^2) + 
\frac{1}{3}\, m_{\pi^0}^2\,,
\ee
to simplify the amplitudes, except in the loops as stated above.
The possible lowest order contributions from the eta mass have been removed
from the amplitudes using the corresponding next-to-leading order relation
as described in \cite{BB}.

The strong LECs,
$L_1^r$ to $L_8^r$, as well as $\sin\epsilon$ 
come from the one-loop fit in \cite{strongiso}, $L_9^r$ from \cite{L9} and
the $G_E$  estimate is from \cite{BP}.

The constant $Z$ from ${\cal L}_{E2}$ we estimate via
\be
Z = \frac{1}{2\,F_\pi^2\,e^2}\, (m_{\pi^+}^2-m_{\pi^0}^2),
\ee
which corresponds to the value in Table~\ref{tab:isoInput}. 
The higher order coefficients of ${\cal L}_{E4}$, $K_1^r\ldots K_{12}^r$,
 are rather unknown.
Some rough estimates exist but we put them to zero here, at the relevant scale.

The IR divergences are canceled by adding the soft-photon Bremsstrahlung.
We have used a 1~MeV cut-off in energy for this and used the same
cut-off in the definition of $F_{\pi^+}$ and $F_{K^+}$. We also use
$m_\gamma=$ 1 MeV, which effectively removes the infrared part.

The subtraction scale $\mu$ is chosen to be $0.77$ GeV unless stated otherwise.

\subsubsection{Input Relevant for the Photon Reducible Diagrams}

The two constants $D_{13}$ and $D_{15}$ are set to zero since no knowledge
exist of their values. One can determine the constants $N_{14}$ and $N_{15}$ 
from \mbox{$K\to\pi l^+l^-$} decays. For a detailed
analysis, see \cite{BB2}. The resulting values are given in 
Table~\ref{tab:isoInput}. Note however, that as described in \cite{BB2}, 
$D_{13}$, $D_{15}$, 
$N_{14}$ and $N_{15}$ only contributes via the photon reducible diagrams. 
These diagrams are negligible numerically, unless the constants are 
orders of magnitude larger than expected.

\subsection{Bremsstrahlung and Dependence on $m_\gamma$ and $\omega$}

The isospin breaking amplitudes for $K\to 2\pi$ and $K\to 3\pi$ both
depend on $m_\gamma$, introduced to regularize the infrared divergences
coming from loops containing photons. This $m_\gamma$-dependence is 
canceled by adding the bremsstrahlung amplitudes, where a real 
soft photon is radiated off one of the mesons. This cancellation was
checked for $K\to 3\pi$ in \cite{BB2}, and we have now also checked it for
$K\to 2\pi$.

However, after the addition of bremsstrahlung the decay rates depend instead
on $\omega$, the cut-off energy of the radiated real photon. This is a 
parameter that should be set to a value depending on the experiment that
one compares to. 

Another possibility, which we use in this paper,
is to add the full amplitudes with an extra radiated 
photon, $K\to 2\pi\gamma$ and $K\to 3\pi\gamma$.
When doing that the decay rates should be independent
of $\omega$. We have checked this numerically and the results are presented
in Table~\ref{tab:omega}. For this comparison we have chosen
$m_\gamma = 0.1$~MeV and varied omega over a large range.
One can see that up to photon energies of 
1 MeV, the sum is constant within the expected uncertainties. The
different sum when including energies up to 10 MeV is an indication
that the soft photon approximation, used in calculating the infrared
contribution, is breaking down.

\begin{table}
\begin{center}
\begin{tabular}{|c|c|c|c|}
\hline
$\omega\, (\mathrm{GeV})$   &   IR photon  &   
Extra photon   &  $\mathrm{Sum}\, 
(\mathrm{GeV})$   \\
   & $(\mathrm{GeV})$  &  $(\mathrm{GeV})$ &  \\
\hline \hline
\multicolumn{4}{|c|}{$K_S\to\pi^+ \pi^-$}\\
\hline
0.01       & $4.34\cdot 10^{-17}$ 
        & $1.59\cdot 10^{-17}$   &  $5.92\cdot 10^{-17}  $                           \\
0.001       &   $2.17\cdot 10^{-17}$ 
        &  $3.69\cdot 10^{-17}$  &  $  5.86\cdot 10^{-17} $                               \\
0.0005       &   $1.52\cdot 10^{-17}$ 
        &  $4.34\cdot 10^{-17}$  &  $  5.85\cdot 10^{-17} $                               \\
0.0001       & $0$ 
        & $5.85\cdot 10^{-17}$   &  $5.85\cdot 10^{-17}$  \\
\hline\multicolumn{4}{|c|}{$K^+\to\pi^+ \pi^0$}\\
\hline
0.01       & $4.30\cdot 10^{-20}$ 
        & $1.54\cdot 10^{-20}$   &  $5.84\cdot 10^{-20}  $                           \\
0.001       &   $2.15\cdot 10^{-20}$ 
        &  $3.61\cdot 10^{-20}$  &  $  5.77\cdot 10^{-20} $                               \\
0.0005       &   $1.50\cdot 10^{-20}$ 
        &  $4.26\cdot 10^{-20}$  &  $  5.76\cdot 10^{-20} $                               \\
0.0001       & $0$ 
        & $5.77\cdot 10^{-20}$   &   $5.77\cdot 10^{-20}$\\
\hline\multicolumn{4}{|c|}{$K_L\to\pi^+ \pi^- \pi^0$}\\
\hline
0.01       & $2.39\cdot 10^{-21}$ 
        & $3.39\cdot 10^{-22}$   &  $2.73\cdot 10^{-21}  $                           \\
0.001       &   $1.19\cdot 10^{-21}$ 
        &  $1.39\cdot 10^{-21}$  &  $  2.58\cdot 10^{-21} $                               \\
0.0005       &   $8.34\cdot 10^{-22}$ 
        &  $1.74\cdot 10^{-21}$  &  $  2.58\cdot 10^{-21} $                               \\
0.0001       & $0$ 
        & $2.56\cdot 10^{-21}$   &  $2.56\cdot 10^{-21}$  \\
\hline
\multicolumn{4}{|c|}{$K_S\to\pi^+ \pi^- \pi^0$}\\
\hline
0.01       & $5.88\cdot 10^{-24}$ 
        & $8.93\cdot 10^{-25}$   &  $6.77\cdot 10^{-24}  $                           \\
0.001       &   $2.94\cdot 10^{-24}$ 
        &  $3.43\cdot 10^{-24}$  &  $  6.37\cdot 10^{-21} $                               \\
0.0005       &   $2.05\cdot 10^{-24}$ 
        &  $4.28\cdot 10^{-24}$  &  $  6.34\cdot 10^{-21} $                               \\
0.0001       & $0$ 
        & $6.29\cdot 10^{-24}$   &  $6.29\cdot 10^{-24}$  \\
\hline
\multicolumn{4}{|c|}{$K^+\to\pi^0 \pi^0 \pi^+$}\\
\hline
0.01       & $3.77\cdot 10^{-22}$ 
        & $4.03\cdot 10^{-23}$   &  $4.18\cdot 10^{-22}  $                           \\
0.001       &   $1.89\cdot 10^{-22}$ 
        &  $1.98\cdot 10^{-22}$  &  $  3.86\cdot 10^{-22} $                               \\
0.0005       &   $1.32\cdot 10^{-22}$ 
        &  $2.53\cdot 10^{-22}$  &  $  3.85\cdot 10^{-22} $                               \\
0.0001       & $0$ 
        & $3.83\cdot 10^{-22}$   &  $3.83\cdot 10^{-22}$  \\
\hline
\multicolumn{4}{|c|}{$K^+\to\pi^+ \pi^+ \pi^-$}\\
\hline
0.01       & $5.05\cdot 10^{-21}$ 
        & $5.86\cdot 10^{-22}$   &  $5.63\cdot 10^{-21}  $                           \\
0.001       &   $2.52\cdot 10^{-21}$ 
        &  $2.73\cdot 10^{-21}$  &  $  5.25\cdot 10^{-21} $                               \\
0.0005       &   $1.76\cdot 10^{-21}$ 
        &  $3.46\cdot 10^{-21}$  &  $  5.23\cdot 10^{-21} $                               \\
0.0001       & $0$ 
        & $5.19\cdot 10^{-21}$   &  $5.19\cdot 10^{-21}$  \\
\hline
\end{tabular}
\end{center}
\caption{$K\to2,3\pi$ decay rates calculated for different values
on $\omega$. Here we use $m_\gamma=$~0.1~MeV and the same value is used
as a cut-off in the decay constants, see the appendix in \cite{BB2}. 
\label{tab:omega}}
\end{table}

The way we treated the Bremsstrahlung contribution in the fits is as follows.
We assume that the measured decay widths are including all photons. To
compare to our amplitudes (calculated without hard photons), we therefore
subtract numerically the calculated hard photon contributions from the
experimental numbers.

\subsection{Fit to $K\to 2\pi$}

The process $K\to2\pi$ in the presence of isospin breaking has
been discussed in detail in \cite{k2piisofull}. We have reproduced that
calculation but added in addition also all the isospin breaking contributions
from the 27 amplitudes, except for the parts from the weak-electromagnetic
27 Lagrangian.

The isospin breaking corrections to the decay rates are rather small,
but they have an impact on the phase shift between the $I=2$ and $I=0$
amplitudes, $\delta_2-\delta_0$,
as described in detail in \cite{k2piisofull}. Our results
are compatible with the ones presented there. The phase shift we use here
is defined via 
\ba
A^S_{00} &=&  \frac{\sqrt2}{\sqrt3} A_0 -\frac{2}{\sqrt3}A_2\,,
\nonumber\\
A^S_{+-} &=& \frac{\sqrt2}{\sqrt3}A_0+\frac{1}{\sqrt3} A_2\,,
\nonumber\\
A_{+0} &=& \frac{\sqrt3}{2} A_2^+\,,
\nonumber\\
\frac{A_2}{A_0} &=& \left|\frac{A_2}{A_0}\right| e^{i(\delta_2-\delta_0)}\,.
\ea
In the fits we have left $\delta_2-\delta_0$ as an additional free parameter,
as it is known that this phase is badly reproduced at one-loop order
in ChPT. Note that because of the isospin breaking the $A_2^+$ amplitude
is different from $A_2$.

We have performed a lowest order and a NLO fit to only the $K\to2\pi$
amplitudes. In the NLO fit we set all the extra parameters, $\tilde K_i =
K_i^r = Z_i^r = 0$ at the scale $\mu$ indicated.
The Bremsstrahlung contribution has been subtracted as discussed above.

\begin{table}
\begin{center}
\begin{tabular}{|c|c|c|c|c|}
\hline
\multicolumn{5}{|c|}{$K\to2\pi$}\\
\hline
Order   & $\mu$ [GeV] & $G_8$ & $G_{27}$ & $\delta_2-\delta_0$\\
\hline
\hline
LO & & 10.4 & 0.55 & $-60.3 \,^\circ$\\
NLO & 0.6 & 6.43 & 0.44 & $-58.9\,^\circ$ \\
NLO & 0.77 & 5.39 & 0.36 & $-58.0\,^\circ$ \\
NLO & 1.0 & 4.60 & 0.30 & $-57.4\,^\circ$ \\
\hline  
\end{tabular}
\end{center}
\caption{Quantities fitted from $K\to2\pi$ only. The scale $\mu$ is the 
scale at which the unknown low energy constants are put to zero.}
\label{tab:K2P}
\end{table}
As can be seen in Table~\ref{tab:K2P}, there is a sizable variation depending
on the input scale used. There is very little change in the
absolute values of $G_8$
and $G_{27}$ compared to the isospin limit fit of \cite{BDP},
where only the fit with $\mu=0.77$~GeV was done. 
The angle is similar to the fit there, but this is a
combination of two different effects. It was 
lowered because the new KLOE data have now
been included in the PDG averaging, but the isospin breaking effects
induced a positive correction as was also found in \cite{k2piisofull}.

The values of $G_8$ and $G_{27}$ are determined by fitting
$C F_0^4 G_8$ and $C F_0^4 G_{27}$ and then setting $F_0=F_\pi$ numerically
to provide the numbers in the tables.

\subsection{Fit to $K\to2\pi$ and $K\to3\pi$}

\begin{table*}
\begin{center}
\begin{tabular}{|c|c|c|c|}
\hline
Decay   & Width [GeV] & ChPT [GeV] & Fact. [GeV]\\
\hline
\hline
$K^+\to\pi^+\pi^0$     &$(1.1231\pm0.0078)\cdot10^{-17}$&$1.123\cdot10^{-17}$
&$1.127\cdot10^{-17}$\\
$K_S\to\pi^0\pi^0$     &$(2.2828\pm0.0104)\cdot10^{-15}$&$2.282\cdot10^{-15}$
&$2.283\cdot10^{-15}$\\
$K_S\to\pi^+\pi^-$     &$(5.0691\pm0.0108)\cdot10^{-15}$&$5.069\cdot10^{-15}$
&$5.069\cdot10^{-15}$\\
\hline
$K_L\to\pi^0\pi^0\pi^0$&$(2.6748\pm0.0358)\cdot10^{-18}$&$2.618\cdot10^{-18}$
&$2.698\cdot10^{-18}$\\
$K_L\to\pi^+\pi^-\pi^0$&$(1.5998\pm0.0271)\cdot10^{-18}$&$1.658\cdot10^{-18}$
&$1.711\cdot10^{-18}$\\
$K^+\to\pi^0\pi^0\pi^+$&$(9.195\pm0.0213)\cdot10^{-19}$&$8.934\cdot10^{-19}$
&$8.816\cdot10^{-19}$\\
$K^+\to\pi^+\pi^+\pi^-$&$(2.9737\pm0.0174)\cdot10^{-18}$&$2.971\cdot10^{-18}$
&$2.933\cdot10^{-18}$\\
\hline  
\end{tabular}
\end{center}
\caption{The various decay widths from the PDG tables \cite{PDG}, 
and our results from the main fit and the best factorization fit.}
\label{tab:widths}
\end{table*}

\begin{table*}[t]
\begin{center}
\begin{tabular}{|c|c|c|c|c|}
\hline
Decay & Quantity & Experiment & ChPT & Fact.\\
\hline
\hline
$K_L\to\pi^0\pi^0\pi^0$ & $h$ & $-0.0050\pm0.0014$  & -0.0062&-0.0025\\
\hline
$K_L\to\pi^+\pi^-\pi^0$ 
    & $g$ & $0.678\pm0.008$   & 0.678 & 0.654\\
    & $h$ & $0.076\pm0.006$   & 0.088 & 0.083\\
    & $k$ & $0.0099\pm0.0015$ & 0.0057 & 0.0068\\
\hline
$K_S\to\pi^+\pi^-\pi^0$ 
 & $\gamma_S$    & $(3.3\pm0.5)\cdot10^{-8}$ & $3.0\cdot10^{-8}$ 
 & $2.9\cdot10^{-8}$\\
\hline
$K^{\pm}\to\pi^0\pi^0\pi^{\pm}$ 
 & $g$ & $0.638\pm0.020$ & 0.636 & 0.648 \\
 & $h$ & $0.051\pm0.013$ & 0.077 & 0.080\\
 & $k$ & $0.004\pm0.007$ & 0.0047 & 0.0069\\
\hline
$K^+\to\pi^+\pi^+\pi^-$ 
 & $g$ & $-0.2154\pm0.0035$ & $-$0.215 & $-$0.226\\
 & $h$ & $0.012\pm0.008$ & 0.012 & 0.019\\
 & $k$ & $-0.0101\pm0.0034$ & $-$0.0034 & $-$0.0033\\
\hline
$K^-\to\pi^-\pi^-\pi^+$ & $g$ & $-0.217\pm0.007$ & &\\
                        & $h$ & $0.010\pm0.006$ & &\\
                        & $k$ & $-0.0084\pm0.0019$ &&\\
\hline
\end{tabular}
\end{center}
\caption{Experimental values and the main fit and best factorization fit 
of the Dalitz plot distribution parameters.
The data are from the PDG tables \cite{PDG} except $\gamma_S$ from
\cite{CPLEAR}.}
\label{tab:dalitz}
\end{table*}
The quantities we fit are the measured decay rates and the various
parameters of the Dalitz plot distributions defined
via
\be
\left|\frac{A(s_1,s_2,s_3)}{A(s_0,s_0,s_0)}\right|^2
= 1 + g y + h y^2 + k x^2\,.
\ee
For the decay $K_L\to3\pi^0$, $k=h/3$ and $g=0$.
The decay $K_s\to\pi^+\pi^-\pi^0$ is included via
\be
A^S_{+-0} = \gamma_S x - \xi xy\,.
\ee

The decay rates are included in the fit as follows. We subtract from the
decay rates the Bremsstrahlung contributions as described above as a function
of $G_8$ and $G_{27}$.
We then convert the decay rate using the central values of the
measured Dalitz plot distribution into a value for the amplitude
squared at the center of the Dalitz plot. These squared amplitudes
together with the parameters $g,h,k$ and $\gamma_S$ are used as the 18
experimental parameters to be fitted. 

This means that the effect of
Bremsstrahlung is included fully in the decay rates, but only via the soft
photon approximation with a 1~MeV cut-off for the Dalitz plot distributions.
We have not included the preliminary
data from KTeV, NA48 and KLOE.

The number of free input parameters on the theory side is very large.
Since it turns out that the isospin breaking effects are very small, we 
put those extra NLO parameters equal to zero at the scale $\mu$
indicated in the tables.

Let us repeat here the definitions of the various extra NLO
parameters. The $L_i^r$ are taken from the standard fit done at one loop
to be compatible with the order of this calculation. The $K_i^r$ are
the extra parameters at NLO in the $p^2 e^2$ sector. Those are always put to
zero at the scale indicated. In the isospin limit 11 combinations of the
weak NLO low-energy coefficients show up, as discussed in \cite{BDP}. These
are $\tilde K_i$, $i=1,\ldots,11$. In the presence of isospin breaking
many more combinations of these, as well as from the weak octet order
$e^2 p^2$ Lagrangian, emerge and they were classified 
in \cite{BB}. The 27-part of the weak
Lagrangian of order $e^2 p^2$ has not been worked out and will lead to some
more free parameters. We have not used any estimates of these extra parameters
but set all of them to zero at the scale indicated, except for
$\tilde K_i$, $i=1,\ldots,7$.
The reason for this choice is that they are the leading contributions.
$\tilde K_{1,2,3}$ are octet enhanced and come multiplied with factors
of order $m_K^4$ and $\tilde K_{4,5,6,7}$ are 27-plets but also 
come multiplied with factors of order $m_K^4$.
The neglected ones are thus suppressed by either isospin breaking,
factors of $m_\pi^2/m_K^2$ or by the $\Delta I=1/2$ rule, i.e.\
an extra factor of $G_{27}/G_8$.

\subsubsection{General Fits}

Here we perform the fits with similar assumptions as used in the isospin limit
fit, as well as a few additional ones. 
First $G_8$ and $G_{27}$ are extremely correlated with the values of
$\tilde K_1$ and $\tilde K_4$ respectively. They are very difficult
to obtain separately without additional assumptions.
The main fit is therefore the one with
\be
\label{constraintmainfit}
\tilde K_1 = \tilde K_4 = \tilde K_8 = \tilde K_9 = 0\,,
\ee
at a scale $\mu=0.77$~GeV. The results are given in Table~\ref{tab:mainfit}.
This table is very similar to Table 6
in \cite{BDP}. The large values of $\tilde K_6$ and the resulting large value
of $\tilde K_7$ have the same origin as in that reference. In order to fit
$\gamma_S$ well, $\tilde K_6$ is put large because it gets multiplied
there with a small factor and is the only one contributing. This in turn
leads large values for $\tilde K_7$ to compensate in other places.

The fit with
\be
\label{constraintK6}
\tilde K_6 = 0
\ee
in addition has only a slightly larger $\chi^2$ and a smaller $\tilde K_7$.
The $\chi^2$ is larger than in \cite{BDP} because the experimental errors
on several quantities have decreased since then. The overall fit is slightly
better than the one of \cite{BDP} because the newer measurements of
the Dalitz distribution in $K^+\to\pi^0\pi^0\pi^+$ agree better with the
chiral expressions.

\begin{table*}
\begin{center}
\begin{tabular}{|c|c|c|c|c|}
\hline
Constraint & Eq. (\ref{constraintmainfit}) & Eq. (\ref{constraintmainfit}) &
 Eq. (\ref{constraintmainfit}) & 
Eq. (\ref{constraintmainfit},\ref{constraintK6}) \\
$\mu$             & 0.77~GeV     & 1.0~GeV      & 0.6~GeV & 0.77~GeV\\
\hline
\hline
$G_8$ &             5.39(1)      & 4.60(1)      & 6.43(1)      &5.39(1)\\
$G_{27}$ &          0.359(2)     & 0.301(1)     & 0.438(2)     &0.359(2)\\
$\delta_2-\delta_0$&$-57.(1.5)^o$&$-57.3(1.4)^o$&$-58.9(1.4)^o$&$-57.9(1.4)^o$
\\
$10^3 \tilde K_1/G_8$ 
& $\equiv 0$   & $\equiv 0$   &  $\equiv 0$  &$\equiv 0$   \\
$10^3 \tilde K_2/G_8$
 & 48.5(2.4)    & 56.5(2.4)    &  41.2(1.9)   &46.6(1.6)\\
$10^3 \tilde K_3/G_8$
 & 2.6(1.2)     & $-$1.7(1.1)  &  6.7(1.0)    &3.5(0.8)\\
$10^3 \tilde K_4/G_{27}$ &
 $\equiv 0$   & $\equiv 0$   &  $\equiv 0$  &$\equiv 0$ \\
$10^3 \tilde K_5/G_{27}$  &
$-$41.2(16.9) &$-$52.0(17.7) &$-$31.1(12.0) &$-$27.0(8.3)\\
$10^3 \tilde K_6/G_{27}$ & 
$-$102(105)   &$-$114(105)   &$-93(76)$     & $\equiv 0$ \\ 
$10^3 \tilde K_7/G_{27}$ &
78.6(33)      &78.0(33.5)    &79.6(22.7)    & 50.0(13.0)\\
$10^3 \tilde K_8/G_8$ &
 $\equiv 0$   & $\equiv 0$   &  $\equiv 0$  &$\equiv 0$   \\
$10^3 \tilde K_9/G_8$ &
 $\equiv 0$   & $\equiv 0$   &  $\equiv 0$  &$\equiv 0$ \\
\hline
$\chi^2$/DOF      & 29.3/10      & 27.2/10      & 33.0/10      &30.5/11\\
\hline
\end{tabular}
\end{center}
\caption{\label{tab:mainfit} The results for $G_8$ and $G_{27}$ and the
$\tilde K_i$ for the various constraints described in the text. In
brackets are the MINUIT errors. The $\tilde K_i$ are quoted at the scale $\mu$
mentioned.}
\end{table*}

We get fits of roughly similar quality for all values of $\mu$ where the
other parameters have been put to zero. The fits tend to be slightly
better for the larger values of $\mu$. The fitted values for the $\tilde K_i$
are $\mu$-dependent, albeit not extremely strongly.

The $\tilde K_i$ themselves have a $\mu$-dependence which is given by
the cancellation of divergences, and this can be calculated from the known
subtractions. We have shown the variation with $\mu$ from $\mu=0.77~GeV$ to 
$\mu=0.6$~GeV and $\mu=1.0$~GeV for $\tilde K_i$, $i=1,\ldots,11$
in Table~\ref{tab:mu}

In order to compare with the factorization model of the weak low energy
constants, we also perform a fit where all next-to-leading order 
LECs proportional to $G_{27}$ are set to zero,
but we keep in addition the sub-leading octet ones. This fit is shown
for $\mu=0.77$~GeV in Table~\ref{tab:mu}. The fit is somewhat worse
than those of Tab.~\ref{tab:mainfit} but not much.
A very similar fit is obtained for $\mu=1$~GeV with a $\chi^2$
of 29.9. At $\mu=0.6~GeV$ the best solution found had a $\chi^2$ of 57.8. 
This fit corresponds to
\be
\label{constraintoctet}
\tilde K_4 =\tilde K_5 =\tilde K_6 =\tilde K_7 = 0\,,
\ee
and this type of fit is referred to below as an octet fit.

\begin{table}
\begin{center}
\begin{tabular}{|c|c|c|c|}
\hline
Constraint &  Eq. (\ref{constraintoctet}) & $\mu$ variation& $\mu$ variation\\
$\mu$      & 0.77~GeV     & 1.0~GeV      & 0.6~GeV \\
\hline
\hline
$G_8$ &             4.84(1)      & -- & -- \\
$G_{27}$ &          0.430(1)     & -- & -- \\
$\delta_2-\delta_0$&$-57.9(0.2)^o$& -- & -- \\
$10^3 \tilde K_1/G_8$ 
& 2.0(1)   &$-$5.88  & 5.61\\
$10^3 \tilde K_2/G_8$
 & 63.0(1.5)    &$-$2.69 &2.57\\
$10^3 \tilde K_3/G_8$
 & $-$6.0(7)   &0.159 & $-$0.152\\
$10^3 \tilde K_4/G_{27}$ &
 $\equiv 0$   & $-$9.93& 9.48\\
$10^3 \tilde K_5/G_{27}$  &
 $\equiv 0$ &0 & 0\\
$10^3 \tilde K_6/G_{27}$ & 
 $\equiv 0$ &27.0 & $-$25.8\\
$10^3 \tilde K_7/G_{27}$ &
 $\equiv 0$ &$-$21.5 & 20.5\\
$10^3 \tilde K_8/G_8$ &
 20.4(1) &$-$0.546 & 0.521\\
$10^3 \tilde K_9/G_8$ &
 9.1(1) &$-$2.92 & 2.79\\
$10^3 \tilde K_{10}/G_8$ &
 $\equiv 0$ &11.6& $-$11.1\\
$10^3 \tilde K_{11}/G_8$ &
 $\equiv 0$ &$-$1.66 & 1.58\\
\hline
$\chi^2$/DOF      & 33.3/10      & -- & -- \\
\hline
\end{tabular}
\end{center}
\caption{\label{tab:mu} The results for $G_8$ and $G_{27}$ and the
$\tilde K_i$ for the octet constraint described in the text. In
brackets are the MINUIT errors. 
The $\tilde K_i$ are quoted at the scale $\mu$
mentioned. The last two columns are the values of the $\tilde K_i$
at the scale $\mu$ mentioned when they are zero at $\mu=0.77~GeV$
and run with $G_8=5.39$ and $G_{27}=0.359$.}
\end{table}

\subsubsection{Fits to Factorization and other Models}

Various models for the NLO weak constants exist. We will discuss here
some of the ones which are presented in \cite{EKW}. These have been
discussed in that reference only for the pure octet case. So the quality
of the models should be compared with the octet fit above.

A first choice is the resonance exchange domination of the weak constants.
The problem here is that the weak decays of the resonances involve themselves
many new unmeasured parameters and thus leads to fairly 
few general conclusions.
If we assume that the vector octet exchange dominates, we get a relation
between the octet NLO constants
\be
N_1^r+N_2^r+2N_3^r = 0\,,
\ee
which is a combination we can in fact determine. It translates for our
parameters into
\be
\tilde K_3 = -\frac{1}{2}\tilde K_2\,.
\ee
It can be easily seen from Tables~\ref{tab:mainfit} and \ref{tab:mu} that
this is very far from being satisfied by our fits.

A very often used model is the factorization model. It corresponds to taking
the underlying four quark operator and bosonizing separately the two
quark currents present there. Looking at the dominant octet operators only
for the cases that we need here, this leads to the relations \cite{EKW}
\ba
N_1^r & =& 2   k_f (32 /3 \, L_1^r+4 \,  L_3^r+2 /3 \,  L_9^r)\,,\nonumber\\
  N_2^r &=& 2  k_f (16 /3  L_1^r+4  L_3^r+10 /3  \, L_9^r)\,,\nonumber\\
  N_3^r &=& 2  k_f (8  L_2^r-2  L_9^r)\,,\nonumber\\
  N_4^r &=& 2  k_f (-16 /3  \, L_1^r-8 /3 \,  L_3^r-4 /3 \,  L_9^r)\,,\nonumber\\
  N_5^r &=& 2  k_f (-L_5^r)\,,\nonumber\\
  N_6^r &=& 2  k_f (2 /3  \, L_5^r)\,,\nonumber\\
  N_7^r &=& 2  k_f (L_5^r)\,,\nonumber\\
  N_8^r &=& 2  k_f (4  \, L_4^r+2  \, L_5^r)\,,\nonumber\\
  N_9^r &=& N_{10}^r = N_{11}^r =N_{12}^r = N_{13}^r = 0\,.
\ea
The parameter $k_f$ allows for some overall adjustment. The special value
$k_f= 1/2$ is referred to as the Weak Deformation Model (WDM)\cite{EKW,EPR}.
We have performed a fit leaving both $k_f$ and $\mu$ free.
Note that the scale $\mu$ is also the scale where we have put all the other
NLO parameters equal to zero. The input values of the $L_i^r$ have been
scaled accordingly.

The fits done with $k_f$, $G_8$ and $G_{27}$ as free parameters have
$\chi^2$ significantly larger than those reported above. Some representative
values are shown in Table~\ref{tab:factorization}. 

The fit with $\mu$ free
gave a minimum at $\mu=0.842$~GeV. The fits with $\mu$ outside the range
of Table~\ref{tab:factorization} had very large values of $\chi^2$.
\begin{table}
\begin{center}
\begin{tabular}{|c|c|c|c|}
\hline
$\mu$    &  0.77~GeV & 0.9~GeV  & 0.842~GeV \\
\hline
\hline
$G_8$    & 4.18(1)   & 4.42     & 4.22(1)   \\
$G_{27}$ & 0.360(2)  & 0.326(10)& 0.339(10) \\
$k_F$    & 2.61(1)   & 4.94(2)  & 3.60(5)   \\
\hline
$\chi^2$/DOF& 109/14 & 182/14   & 60.4/13   \\
\hline
\end{tabular}
\end{center}
\caption{\label{tab:factorization} The results for the fit with the
factorization assumption for various values of $\mu$ including the
optimal one.}
\end{table}

In order to show the quality of the fits we have given in
Tables~\ref{tab:widths} and \ref{tab:dalitz} also the values obtained for the
quantities from the main fit and best factorization fit, labeled
ChPT and Fact.\ respectively. Notice that the extrapolation
to the full phase space here has been done from the amplitude squared
in the center of the Dalitz plot using the experimental values for
the distribution over the Dalitz plot.

\section{Summary}
\label{summary}

We have recalculated in this paper the Bremsstrahlung amplitudes
for $K\to2\pi\gamma$ and $K\to3\pi\gamma$.
In addition we have calculated also the isospin violating effects
to $K\to2\pi$ including those with the 27-operators both for
effects due to $m_u-m_d$ and electromagnetism.
 This we did to be consistent with the calculations of $K\to3\pi$
done in \cite{BB,BB2}.

We have checked explicitly that the infrared divergences of the photon
loops regulated by $m_\gamma$ cancel between the virtual photon loops
and the soft Brems-strahlung. We checked in addition that the photon energy
cut-off dependence cancels between the soft-photon Bremsstrahlung part
and the part where hard photons are treated explicitly. 
We have not included the explicit expressions for the $K\to2\pi$ amplitudes 
because they are 
rather long. They can be obtained from the
authors or \cite{formulas}. These amplitudes have passed all the standard
tests, like cancellation of divergences from both NLO ChPT as well as the 
infrared singularities.

With these calculations and those published earlier in \cite{BB,BB2}, we have
updated the fit to the CP conserving observables in the
$K\to \pi\pi(\pi)(\gamma)$ system done in the isospin limit in \cite{BDP}.
As expected from the fairly small isospin violating effects found in
\cite{BB,BB2} and from the analysis of isospin breaking effects in the
$K\to\pi\pi(\gamma)$ system of \cite{k2piisofull}, the differences with
the isospin conserving case are rather small. In addition we have studied
the dependences on the subtraction scale $\mu$, where the various assumptions
are made. Our full amplitudes are $\mu$ independent as they should.

The fits show a similar quality to the ones performed earlier. The main
differences are that the experimental Dalitz parameters have changed in 
$K^+\to\pi^+\pi^+\pi^0$ and are now in better agreement with the ChPT
fits. This is purely experimental and has nothing to do with the inclusion
of isospin breaking effects. The total $\chi^2$ is somewhat worse because the
experimental errors on various partial widths have been reduced.

We also checked how well a few models of the NLO weak low energy constants
work. The dominance by vectors and the weak deformation model gave 
a rather bad fit.
The factorization model gave a somewhat better fit when an extra parameter,
an overall scale factor, was allowed. The quality of this compared
to the optimal ChPT fits can be judged from the tables giving the best fit
values for the experimental quantities directly.

\section*{Acknowledgments}
The program FORM 3.0 has been used extensively in these calculations
\cite{FORM}. This work is supported in part by the Swedish Research Council
and European Union TMR
network, Contract No. HPRN-CT-2002-00311  (EURIDICE).

\end{document}